\documentclass{article}%
\usepackage{amsmath}%
\usepackage{amsfonts}%
\usepackage{amssymb}%
\usepackage{graphicx}

\ifx\pdfoutput\relax\let\pdfoutput=\undefined\fi
\newcount\msipdfoutput
\ifx\pdfoutput\undefined\else
\ifcase\pdfoutput\else
\ifx\paperwidth\undefined\else
\ifdim\paperheight=0pt\relax\else\pdfpageheight\paperheight\fi
\ifdim\paperwidth=0pt\relax\else\pdfpagewidth\paperwidth\fi
\fi\fi\fi
\begin{document}

\begin{center}
{\LARGE Observation and Quantum Objectivity }
\end{center}

\bigskip

{\large Abstract}

The paradox of Wigner's friend challenges the objectivity of description in
quantum theory. A pragmatist interpretation can meet this challenge by
judicious appeal to decoherence. On this interpretation, quantum theory
provides situated agents with resources for predicting and explaining what
happens in the physical world---not conscious observations of it. Even in
Wigner's friend scenarios, differently situated agents agree on the objective
content of statements about the values of physical magnitudes. In more
realistic circumstances quantum Darwinism also permits differently situated
agents equal observational access to evaluate their truth. In this view,
quantum theory has nothing to say about consciousness or conscious experiences
of observers. But it does prompt us to reexamine the significance even of
everyday claims about the physical world.

\bigskip

\section{Introduction}

Wigner (1967) presented his "friend paradox" to motivate the view that
consciousness has a special role to play in quantum theory---by collapsing the
quantum state. In a more recent discussion, D'Espagnat (2005) argued that a
consistent treatment of the Wigner's friend scenario may be given (without
collapse) if the descriptive content underlying quantum theory is restricted
to probabilistic predictions flowing from the Born Rule, but only if these
concern conscious experiences of observers. In contrast, a recent pragmatist
proposal (Healey 2012) views quantum theory as a resource for situated
\textit{agents} (rather than observers) with no requirement that these be
conscious.\footnote{As far as we know, conscious humans are the only agents
currently able to avail themselves of this resource. But this view admits the
possibility that non-human, or even non-conscious, situated agents may come to
use quantum theory.} If quantum theory is interpreted along these pragmatist
lines, Wigner's friend scenarios may be treated consistently and without ambiguity.

The paper proceeds as follows. Section 2 analyzes distinct notions of
objectivity relevant to quantum theory. Section 3 recalls the paradox of
Wigner's friend. The pragmatist interpretation of Healey (2012) is sketched in
section 4. Section 5 shows how this resolves the paradox while securing the
objectivity of description in quantum theory. Section 6 addresses a more
general worry about objectivity of description. Section 7 relates the
foregoing treatment of objectivity to a recent suggestion (Ollivier \textit{et
al}. 2004) that objective properties emerge from subjective quantum states
through quantum Darwinism. While a pragmatist view of quantum theory secures
the objectivity of a claim about the values of physical magnitudes, it
recognizes that its content depends on the environmental context. The
conclusion points out that this involves a subtle change in our understanding
of how even the familiar language of everyday affairs, including laboratory
procedures, assigns objective properties--beables--to objects (Bell 2004).

\section{Objectivity}

Objectivity (and its polar opposite, subjectivity) can mean many things.
Dreams are a paradigm of subjectivity. The subject matter of a dream is not
objective: a dream does not portray what really happens. The mode of
presentation is not objective: this subject matter is accessible only to the
dreamer. What \ is presented as well as its mode of presentation strongly
depends on the specific as well as the general features of the dreamer---her
individual physiology, psychology and prior experiences as well as her
humanity (Dick 1968). By contrast, a description or mathematical
representation in physics has an objective subject matter if its content
represents physical reality, and an objective mode of presentation to the
extent that how this is represented does not depend on the specific and
general features of the one whose representation it is.

The prominence of notions of observation or measurement in standard
formulations of quantum theory raises concerns about the objectivity of
descriptions in that theory. If the Born Rule is understood to yield
probabilities just for \textit{results of observations/measurements} then one
can question the objectivity of these results. The orthodox view---that a
quantum measurement cannot be understood generally as revealing the value of
the measured observable---has now been amply supported by "no-go" theorems
(Gleason 1957, Kochen and Specker 1967, Fine 1982, Mermin 1993). This
challenges the objectivity of subject matter in a description of the result of
an observation. One can respond to this challenge by proposing an analysis of
observation/measurement as an objective physical process. But the assumption
that quantum theory can itself represent this process leads to the notorious
quantum measurement problem (Fine 1970, Brown 1986, Bassi and Ghirardi 2000).
The alternative of characterizing observation independently as a physical
interaction with a macroscopic apparatus, and/or involving irreversible
amplification involves the "shifty split" between quantum and classical
descriptions justifiably criticized in Bell (2004).

With his keen appreciation of how difficult it is to provide a satisfactory
physical characterization of observation/measurement in quantum theory, Wigner
came to think that a measurement in quantum theory occurs just when a
conscious observer becomes aware of a result. On this view, the Born Rule
yields probabilities only for alternative conscious experiences. It thereby
threatens both the objectivity of the subject matter of description in quantum
theory and the objectivity of its mode of presentation. Quantum theory, it
seems, is then concerned to predict and (perhaps) explain "communicable human
experience. In other words...the set of all the impressions human minds may
have and communicate to others." (D'Espagnat 2005)\footnote{The Wigner's
friend scenario casts doubt even on such communicability. D'Espagnat's paper
seeks to relieve this doubt.}

Suppose one so restricts the scope of quantum theory. Then the descriptive
claims to which the Born Rule attaches probabilities fail to be objective in
several respects:

(1) Their subject matter is not a physical reality independent of our experiences.

(2) Their mode of presentation depends on the individual consciousness.

(3) This consciousness is generically \textit{human.}\newline One could seek
to restore this third respect of objectivity by generalizing beyond humans to
observers capable of any sufficiently sophisticated form of conscious
experience. But since consciousness is not well understood (especially in
non-humans) to do so would further obscure the content of these descriptive
claims, which would still fail to be objective in either of the first two respects.

While the quantum theory of environmental decoherence does not by itself
resolve the quantum measurement problem (D'Espagnat 1990, Bub 1997, Adler
2000), many believe it may contribute to a resolution within some appropriate
interpretative framework. Such a framework appears in (Healey 2012), and is
sketched in section 4 below. From the present perspective, the key
interpretative proposal is to remove any talk of observation or measurement
from a formulation of the Born Rule by taking this to generate
\textit{mathematical} probability distributions over claims that simply
describe values of magnitudes rather than results of observing them. The
"no-go" theorems block this approach if one further assumes that every claim
over which these mathematical distributions are defined always has a
well-defined content, so an agent should believe it to a degree specified by
its Born "probability" to avoid refutation by statistics of actual
observations. But suppose, on the other hand, that how much significance
attaches to a claim about a system depends on the extent of environmental
decoherence suffered by its quantum state. Then the numbers yielded by the
Born rule have the import of genuine probabilities only for significant
claims: Born rule "probabilities" for claims lacking such significance do not
correspond to (actual or hypothetical) frequencies, and should not guide an
agent's degrees of belief in these insignificant claims.

On this view any reference to observation or measurement has been eliminated
from the Born rule as well as other basic principles of quantum theory, and so
there is no reason to suppose that descriptive claims that arise in quantum
theory are subjective in any of the respects (1)-(3) noted above. But there is
a fourth aspect of objectivity to consider, as the following quote makes clear.

\begin{quote}
A view or form of thought is more objective than another if it relies less on
the specifics of the individual's makeup and position in the world, or on the
character of the particular type of creature that he is. (Nagel 1986)
\end{quote}

If the significance for an agent of a descriptive claim about the value of a
magnitude depends on how that agent is situated in the world, then that claim
may lack a kind of objectivity. Differently positioned agents could understand
the claim differently: they may come to a no-fault disagreement about whether
it is true, or even meaningful. The interpretation of Healey (2012) faces such
a challenge to its understanding of the objectivity of descriptive claims in
quantum theory---a challenge that is highlighted by the Wigner's friend
scenario, as we shall see.

\section{Wigner's friend}

The \textquotedblleft paradox\textquotedblright\ of Wigner's friend presents a
challenge to the objectivity of physical description within quantum theory. To
set up the \textquotedblleft paradox\textquotedblright, imagine
Schr\"{o}dinger's cat (and associated `diabolical device') replaced by a human
experimenter (Wigner's friend, John) who records in a device \textit{D} the
result of a quantum measurement of observable \textit{Q} he performs on a
system \textit{s} inside his isolated laboratory.\footnote{To call the
laboratory `isolated', is to require by fiat the absence of any decohering
interactions with its external environment. So we are talking of a
ridiculously impractical \textit{Gedankenexperiment}, as Schr\"{o}dinger
explicitly said he was when describing his cat scenario. The point of doing so
is to show how this raises a problem for one view of quantum theory and then
to explain why this problem does not arise on the pragmatist view of Healey
(2012).} Eugene and John both agree that the quantum state of \textit{s} is
initially (at time \textit{t}$_{i}$) a non-trivial superposition of
eigenstates of \textit{Q}. Let \textit{C} be the claim that the value of a
recording magnitude \textit{M} on \textit{D} is \textit{m} at time
\textit{t}$_{f}$ after \textit{s} has interacted with \textit{D}. John looks
at \textit{D} at \textit{t}$_{f}$\ and makes claim \textit{C}. On the basis of
this observation, he assigns a "collapsed" quantum state to \textit{D}%
+\textit{s}.\footnote{He does so in accordance with a view of quantum states
due to Dirac and von Neumann. While himself subscribing to this orthodoxy,
Wigner maintained that it is only John's consciousness that induces the
collapse.} Environmental interactions within the laboratory rapidly entangle
this "collapsed" state with the state of everything else in the laboratory.
But the total quantum state $\left\vert \psi_{J}\right\rangle $\ of the
enormously complex system composed of John, \textit{D} and everything else
inside the laboratory will continue to reflect the "collapse" onto an
eigenstate of \textit{M} induced by John's measurement.

Meanwhile Eugene, who has remained outside the laboratory, assigns a state
$\left\vert \psi_{E}\right\rangle $ to the enormously complex system composed
of John, \textit{D} and everything else inside the laboratory, based on all
the information about the properties of systems to which he has access in this
situation. For\ Eugene, John's measurement involves purely unitary
interactions---between \textit{s} and \textit{D}, between \textit{D} and John,
and between all these systems and the rest of the laboratory. Accordingly, he
evolves the state $\left\vert \psi_{E}\right\rangle $ unitarily, with no
"collapse" from \textit{t}$_{i}$ to \textit{t}$_{f}$. Since $\left\vert
\psi_{J}(t_{f})\right\rangle $\ $\neq$ $\left\vert \psi_{E}(t_{f}%
)\right\rangle $, John's and Eugene's quantum states differ after John's
measurement but before Eugene enters the laboratory. As these states are
usually understood, $\left\vert \psi_{J}(t_{f})\right\rangle $ represents a
definite result of John's measurement (recorded by the state of \textit{D},
John's memory, etc.) while $\left\vert \psi_{E}(t_{f})\right\rangle $ excludes
any such definite result and its traces. To retain its internal consistency,
this view of quantum states must deny the objectivity of John's measurement
result, since differently situated agents (John and Eugene) disagree about
whether \textit{C} is true at \textit{t}$_{f}$.

On any view of quantum states, since $\left\vert \psi_{E}(t_{f})\right\rangle
$ is an eigenstate of some extremely complex observable \textit{O }on the
entire laboratory system, Eugene could in principle (though certainly not in
practice) distinguish between $\left\vert \psi_{J}(t_{f})\right\rangle $\ and
$\left\vert \psi_{E}(t_{f})\right\rangle $ by creating a suitable ensemble of
identical laboratory systems and measuring \textit{O} on each system:
$\left\vert \psi_{E}(t_{f})\right\rangle $ would give the same result on each,
while $\left\vert \psi_{J}(t_{f})\right\rangle $\ would almost certainly yield
a statistical spread of different results. This same procedure would also, in
principle, enable Eugene to distinguish between $\left\vert \psi_{E}%
(t_{f})\right\rangle $ and the mixed state $\rho_{J}(t_{f})=%
{\textstyle\sum_{i}}
\left\vert c_{i}\right\vert ^{2}\left\vert \psi_{J}(t_{f})\right\rangle
_{i}\left\langle \psi_{J}(t_{f})\right\vert _{i}$ Eugene may assign to reflect
his ignorance of the result of John's measurement.\footnote{To keep things
simple, here I assume John made an ideal measurement on the state $%
{\textstyle\sum_{i}}
c_{i}\left\vert \varphi_{i}\right\rangle $, thereby collapsing the state of
\textit{s}+\textit{D} onto some eigenstate $\left\vert \varphi_{j}%
\right\rangle \left\vert \psi_{j}\right\rangle $ with probability $\left\vert
c_{j}\right\vert ^{2}$.}

Wigner's own way of resolving this paradox was to give consciousness (and only
consciousness) the distinctive physical role of inducing "collapse" of the
quantum state onto an eigenstate of the measured observable. For him, it was
the interaction with John's consciousness that produced a discontinuous
physical change inside the laboratory, resulting in the final state
$\left\vert \psi_{J}(t_{f})\right\rangle $\ and not $\left\vert \psi_{E}%
(t_{f})\right\rangle $. Such a change would be detectable in principle in the
Wigner's friend scenario, though quite impossible to detect in practice. I
will offer a different resolution involving no such physical "collapse". This
involves the different understanding of quantum states described in the next section.

\section{A pragmatist interpretation}

In quantum theory as it is usually formulated, theoretical models involve
quantum states and operators corresponding to observables (including the
Hamiltonian and/or Lagrangian) and (solutions to) the Schr\"{o}dinger equation
and relativistic generalizations. But there is still no agreement as to how,
or whether, any of these model elements represent physical magnitudes.

\qquad According to Healey (2012), observables, quantum states and their
evolution neither represent nor describe the condition or behavior of any
physical system. It follows that quantum theory does not imply statements one
can use to make claims about natural phenomena that describe or represent
features of those phenomena. But quantum theory nevertheless helps us to
predict and explain an extraordinary variety of physical phenomena using
representational resources from outside of quantum theory. It can do this
because of the way in which theoretical models involving quantum states,
operators, and the Schr\"{o}dinger equation are applied. There are principles
for using these models to guide one in making descriptive claims and forming
representational beliefs about physical systems to which such models may be
applied but which these models do not themselves describe or represent.

\qquad The dispute as to whether quantum theory provides a complete
description of a physical system presupposes that quantum states at least
provide a partial description or representation of the physical properties of
systems to which they are assigned. Rejecting this presupposition may seem
tantamount to regarding the quantum state as merely a symbolic device for
calculating probabilities of possible measurement outcomes on these systems.
But this is not so. Assignment of a quantum state may be viewed as merely the
first step in a procedure that licenses a user of quantum theory to express
claims about physical systems in descriptive language and then warrants that
user in adopting appropriate epistemic attitudes toward these claims. The
language in which such claims are expressed is not the language of quantum
states or operators, and the claims are not about probabilities or measurement
results: they are about the values of magnitudes. That is why Healey (2012)
refers to such claims as NQMC's---Non-Quantum Magnitude Claims. Here are some
typical examples of NQMC's:

\begin{quote}
A helium atom with energy $-24.6$ electron volts has zero angular momentum.

Silver atoms emerging from a Stern-Gerlach device each have angular-momentum
component either +$\hbar$/2 or $-\hbar$/2 in the \textit{z}-direction.

The fourth photon will strike the left-hand side of the screen.

When a constant voltage \textit{V} is applied across a Josephson junction, an
alternating current\textit{\ I} with frequency 2(\textit{e}/\textit{h}%
)\textit{V} flows across the junction.
\end{quote}

(Notice that two of these non-quantum claims are stated in terms of Planck's
constant.) For contrast, here are some quantum claims which do not describe
the physical properties of systems to which they pertain:

\begin{quote}
The expectation value of angular momentum for atomic helium in the ground
state is 0.

Integral-spin systems have symmetric quantum states, while half-integral-spin
systems have antisymmetric quantum states.

The probability that a tritium nucleus will decay in 12.3 years is
${\frac12}$%
.

After one photon from the polarization-entangled Bell state
$\vert$%
$\Phi$+%
$>$
is found to be horizontally polarized, the other photon has polarization
state
$\vert$%
H%
$>$%
.
\end{quote}

This is not to say that quantum mechanical claims like these lack
truth-values---each is appropriately evaluated as objectively true (though in
the case of the last claim that evaluation is critically dependent on the
context relative to which it is made). But the function of such claims is not
to describe or represent properties of physical systems: it is to offer
authoritative advice to a physically situated agent on the content and
credibility of NQMC's concerning them. Quantum theory contributes indirectly
to our predictive and explanatory projects.

\qquad A quantum state and the consequent Born probabilities can be assigned
to a system only relative to the physical situation of an (actual or
hypothetical) agent for whom these assignments would yield good epistemic
advice. What one agent should believe may be quite different from what another
agent in a different physical, and therefore epistemic, situation should find
credible. This relational character of quantum states and Born probabilities
does not make these subjective, and it may be neglected whenever users of
quantum theory find themselves in relevantly similar physical situations.
NQMC's are not relational in this way: Their truth-values do not depend on the
physical situation of any actual or hypothetical agent.

\qquad On this pragmatist understanding, a quantum state guides an agent in
two different ways. The agent requires guidance in assessing the content of
NQMC's about systems of interest in a context where such claims may arise. It
is often said that assignment of a value to an observable on a system is
meaningful only in the presence of some apparatus capable of measuring the
value of that observable. But some account of meaning must be offered in
support of this assertion, and the extreme operationist account that is most
naturally associated with it would be unacceptably vague even if it were
otherwise defensible. Exactly what counts as the presence of an apparatus
capable of measuring the value of an observable?

\qquad Contemporary pragmatist approaches to meaning have the resources to
provide a better account of the significance of a NQMC about a system, as
entertained by an agent, in a context in which that system features. A
pragmatist like Brandom (1994, 2000) takes the content of any claim to be
articulated by the material inferences (practical as well as theoretical) in
which it may figure as premise or conclusion. These inferences may vary with
the context in which a claim arises, so the content of the claim depends on
that context. The quantum state modulates the content of NQMC's about a system
by specifying the context in which they arise. The context may be specified by
the nature and degree of environmental decoherence suffered by this quantum
state. A NQMC about a system when the quantum state has extensively decohered
in a basis of eigenstates of the operator corresponding to that magnitude on
the system has a correspondingly well-defined meaning: a rich content accrues
to it via the large variety of material inferences that may legitimately be
drawn to and from the NQMC in that context. Call a NQMC canonical if it is of
the form $M\in\Delta$:

\begin{quote}
Magnitude $M$ has value in Borel set $\Delta$ of real numbers.
\end{quote}

Only when the content of a canonical NQMC is sufficiently well articulated in
this way is it appropriate to apply the Born Rule to assign a probability to
that claim.

For example, a claim that an electron passed unobservably through a particular
slit in a diffraction grating figures as premise or conclusion in almost no
interesting material inferences, and so is very poorly articulated. This is a
consequence of the fact that, in the absence of interactions capable of
revealing its presence, the quantum state of the electron suffers negligible
decoherence through entanglement with the environment in the interferometer.
With no significant decoherence between different spatially localized quantum
states, an agent has only a very limited license to use claims about the
electron's position. In particular, the license is so limited that little may
legitimately be inferred from such claims. But the subsequent interaction with
detectors at the screen involves massive environmentally induced decoherence
of the initial quantum state of electron plus detector, permitting a high
degree of articulation of the content of claims about the value of a magnitude
on the detector taken to record the position of the electron at the screen.
The decohered quantum state then grants an agent a wide license to formulate
and use NQMC's about records of the electron's position at the screen. This
illustrates the first way in which a quantum state guides an agent---by
advising her on the extent of her license to use particular NQMC's by
informing her of the nature and degree of that state's environmental decoherence.

\qquad With a sufficiently extended license, an agent may now apply the Born
Rule to evaluate the probability of each licensed canonical NQMC using the
appropriate quantum state. In the example, this will be the initial superposed
state of the electron. Lacking more direct observational information, after
assigning the quantum state of a system appropriate to her physical situation,
an agent should adjust her degrees of belief in licensed NQMC's pertaining to
that system so they match the probabilities of NQMC's specified by the Born
Rule. This is the second way in which a quantum state guides an agent.

\qquad An agent should accept this twofold guidance by the quantum state
appropriate for one in her physical situation, since by doing so she is able
successfully to predict and explain what happens in a wide variety of
circumstances. But according to Healey (2012) she need not, and should not,
take this quantum state to describe or represent ontological features of a
system to which she ascribes it. The interpretation does not deny that
environmental decoherence involves a physical process. But it does deny that
decoherence of the quantum state represents changes in intrinsic properties of
a physical system. An agent appeals to the quantum theory of decoherence to
decide what to think about the content and credibility of NQMCs about a
physical system, not to describe the evolving properties of a system whose
quantum state decoheres.

\section{Paradox resolved}

Consider the situation of Wigner's friend John inside his isolated laboratory.
According to the pragmatist interpretation sketched in the previous section,
John is licensed by decoherence to apply the Born Rule to the superposed state
of \textit{s} and advised by quantum theory to adjust his credences in claims
about the value of recording magnitude \textit{M} on \textit{D} at time
\textit{t}$_{f}$ so they match the corresponding Born probabilities. This is
because the phase of the quantum state he consequently assigned to
\textit{D}+\textit{s} immediately before has, by \textit{t}$_{f}$, been
robustly delocalized into the environment inside his laboratory. He is then
warranted by his own direct observation of \textit{D} in making the
significant claim \textit{C} that the value of \textit{M} on \textit{D} is
\textit{m}.

Meanwhile Eugene, who has remained outside the laboratory, assigns state
$\left\vert \psi_{E}\right\rangle $ to the enormously complex system composed
of John, \textit{D} and everything else in the laboratory. By assumption, the
phase of state $\left\vert \psi_{E}\right\rangle $ has not been delocalized
into its environment by \textit{t}$_{f}$, so this state does not license
Eugene to entertain any significant claim about the value of a magnitude on
the laboratory or anything in it. In this situation, Eugene should not apply
the Born Rule to $\left\vert \psi_{E}\right\rangle $ and adjust any of his
credences accordingly. If and only if the laboratory ceases to be isolated so
as to delocalize the phase of $\left\vert \psi_{E}\right\rangle $ (as Eugene
subsequently enters it, for example) would Eugene be in a position to apply
the Born Rule to significant claims about the laboratory or traces it leaves
in an external recording device (such as his own brain).

Recall the difficulties presented by the Wigner's friend scenario noted in
section 3:

1) As these states are usually understood, $\left\vert \psi_{J}(t_{f}%
)\right\rangle $ represents a definite result of John's measurement (recorded
by the state of \textit{D}, John's memory, etc.) while $\left\vert \psi
_{E}(t_{f})\right\rangle $ excludes any such definite result and traces of it
in John's laboratory.

2) On any view of quantum states, since $\left\vert \psi_{E}(t_{f}%
)\right\rangle $ is an eigenstate of some extremely complex observable
\textit{O }on the entire laboratory system, Eugene could in principle (though
certainly not in practice) distinguish between\ $\left\vert \psi_{E}%
(t_{f})\right\rangle $ and $\left\vert \psi_{J}(t_{f})\right\rangle $ (or
$\rho_{J}(t_{f})$)\ by creating a suitable ensemble of identical laboratory
systems and measuring \textit{O} on each system.

These difficulties present a challenge to the objectivity of physical
description here. (1) apparently implies that, by assigning different quantum
states to the contents of John's laboratory at $t_{f}$, John and Eugene come
to disagree about the truth-value of the descriptive claim \textit{C}. If so,
they can both be right only if that truth-value is relative to the physical
situation of the agent making it---in conflict with the fourth aspect of
objectivity noted in section 2. But (2) shows that such relativization of
truth-value to agent-situation still leaves it unclear how John and Eugene can
consistently apply quantum theory here. Independent of the practicality of
Eugene's discriminatory measurements, $\left\vert \psi_{E}(t_{f})\right\rangle
$ and $\left\vert \psi_{J}(t_{f})\right\rangle $ (or $\rho_{J}(t_{f})$) are
distinct states, yielding incompatible Born probabilities concerning the
possible values of certain magnitudes on the entire laboratory. Relativization
of the laboratory's quantum state to the situations of Eugene, John,
respectively leaves it ambiguous on which of these states quantum theory
advises them to base their expectations about these possible values.

The pragmatist of the previous section responds to this challenge, first by
rejecting the usual understanding of the states $\left\vert \psi_{J}%
(t_{f})\right\rangle $, $\left\vert \psi_{E}(t_{f})\right\rangle $. For her,
neither state has the function of representing the physical properties of
John's laboratory or anything in it. Since neither state represents anything
bearing on the result of John's measurement (e.g. whether or not \textit{C} is
true at $t_{f}$), assignment of both states to John's laboratory at the same
time could not lead John and Eugene to assign (apparently) conflicting
truth-values to \textit{C.}

But John and Eugene each assume that John will perform a measurement on
\textit{s} by interacting \textit{s} appropriately with \textit{D}, and that
the environment inside the laboratory will decohere the state of
\textit{D}+\textit{s} in \textit{D}'s "pointer basis" (without inducing any
physical collapse in that state). So both John \textit{and Eugene} are
licensed to entertain claim \textit{C} and (beforehand) to set their credences
for \textit{C} at $t_{f}$ equal to the corresponding Born Rule probability.
This implies they agree that \textit{C} has a truth-value at $t_{f}$. But
while Eugene remains outside the laboratory, only John is in a position to
look at \textit{D} at and after $t_{f}$ and so determine what that truth-value
is. This secures the objectivity of the description \textit{C}, in the sense
that differently positioned agents (John and Eugene, in this case) agree that
\textit{C} has a truth-value after $t_{f}$, and do not disagree about what
that truth-value is. At this stage, John, but not Eugene, is in a position to
\textit{know} that \textit{C} is true rather than false. Eugene can choose
whether or not to enter the laboratory to try to find out whether \textit{C}
is true.

Suppose Eugene decides to see for himself the outcome of John's observation by
entering the laboratory just after $t_{f}$. This will decohere the phase of
$\left\vert \psi_{E}(t_{f})\right\rangle $, permitting Eugene to apply the
Born Rule to this state to adjust his degrees of belief in various significant
claims about the contents of the laboratory, including \textit{C}, John's
record of the outcome in his notebook, John's verbal report, etc. as well as
correlations between such claims. The Born probabilities of these claims
(joint as well as single) \ based on the state $\left\vert \psi_{E}%
(t_{f})\right\rangle $ will lead him to the following confident expectation:
whatever may be the outcome of John's observation, the laboratory will contain
multiple mutually supporting records of it. Nothing about $\left\vert \psi
_{E}(t_{f})\right\rangle $ will tell him what John's outcome actually was. But
he can easily find that out by asking John and observing any of the other
multiple, correlated records inside the laboratory. The important point is
that $\left\vert \psi_{E}(t_{f})\right\rangle $ does not \textit{exclude} an
outcome at $t_{f}$, so by entering the laboratory Eugene simply finds out what
happened---he does not \textit{make} it happen.

If Eugene were instead (able) to remain outside the laboratory but interact
with it so as to measure the value of \textit{O}, he should use $\left\vert
\psi_{E}(t_{f})\right\rangle $ to calculate a Born probability of $1$ that a
suitable recording device, applied to the laboratory and decohered in its
"pointer basis" by external environmental interactions, will record a value of
\textit{O} equal to the eigenvalue of its corresponding operator in state
$\left\vert \psi_{E}(t_{f})\right\rangle $. Even though Eugene assumes that
John records the outcome of his measurement at $t_{f}$, by $t_{f}$\ there has
been no physical interaction between Eugene and anything inside the laboratory
that could serve to inform Eugene of that outcome. So Eugene cannot base his
expectation of the outcome of a measurement of \textit{O} on $\left\vert
\psi_{J}(t_{f})\right\rangle $, and would be mistaken if he were to base that
expectation on $\rho_{J}(t_{f})$. There is no ambiguity about to what state
Eugene should apply the Born Rule when setting his credences concerning the
outcome of his \textit{O} measurement.

On what quantum state should John base his credence as to the outcome of an
external \textit{O}-measurement? The outcome of John's measurement should
prompt him to update his quantum state for objects in his laboratory. These
include \textit{D}, \textit{s} and other systems inside the laboratory, but
\textit{not} himself, considered as a physical system: the \ function of
John's quantum state is to set his credences concerning claims about these
other systems, but not about himself. John should not\ base his credences as
to the outcome of an external \textit{O}-measurement on either of the states
$\left\vert \psi_{J}(t_{f})\right\rangle $ or $\rho_{J}(t_{f})$, since these
are states \textit{of himself }as well as everything else in his laboratory.
He can hypothetically take the "external" perspective\ of Eugene, but if he
does so he must then assign the same state $\left\vert \psi_{E}(t_{f}%
)\right\rangle $ that he agrees is the correct state for someone in Eugene's
situation. There is no ambiguity as to which quantum state should be used to
calculate Born probabilities for the results of external measurements
performed on the entire laboratory, in the sense that differently positioned
agents (John and Eugene, in this case) agree on this state. Of course, only
Eugene is then in a position to make such a measurement.

What will Eugene find if he enters the laboratory after measuring the value of
\textit{O} in this way? \textit{D} will no longer have a record of the truth
of \textit{C}. Even if John still exists, he will not remember that \textit{C}
was verified as true at $t_{f}$. Because $\hat{O}$ fails to commute with the
operator corresponding to \textit{M}, as well as an effective "pointer
observable" on every other system correlated with this through interactions
within the laboratory, the external interaction required to record the value
of \textit{O} will have so disturbed the laboratory and its contents as to
remove all traces that \textit{C} was true at $t_{f}$. But one can't change
the past: \textit{C} was indeed true at $t_{f}$, as John then verified with
his own eyes.

There is, in principle, an even more dramatic way to erase all traces of
\textit{C}. Since the entire laboratory and its contents constitutes an
isolated system, Eugene will take $\left\vert \psi_{E}\right\rangle $ to have
evolved unitarily from its state $\left\vert \psi_{E}(t_{i})\right\rangle $
prior to John's measurement on \textit{s} to its state $\left\vert \psi
_{E}(t_{f})\right\rangle =U_{if}\left\vert \psi_{E}(t_{i})\right\rangle $. The
state Eugene should ascribe to the contents of the laboratory at $t_{i}$
should reflect his belief that his friend has not yet performed the planned
measurement: this will assign Born probability $1$ to claims about \textit{D},
John and other items in the laboratory that suffices to substantiate this full
belief. Mathematically, there will exist a Hamiltonian that would induce the
time-reversed evolution of $\left\vert \psi_{E}\right\rangle $ so that at a
later time $t_{g}$(where $t_{g}-t_{f}=t_{f}-t_{i}$) it is restored to its
value before John's measurement: $\left\vert \psi_{E}(t_{g})\right\rangle
=U_{fg}^{\dag}\left\vert \psi_{E}(t_{f})\right\rangle =\left\vert \psi
_{E}(t_{i})\right\rangle $.

If Eugene had the powers of a quantum demon, he could instantaneously replace
the original Hamiltonian by this time-reversing Hamiltonian at $t_{f} $,
thereby restoring $\left\vert \psi_{E}\right\rangle $ \ at $t_{g}$ to its
original value at $t_{i}$. Suppose that he does so, and postpones his entry
into the laboratory until $t_{g}$. Since the quantum state of the entire
laboratory is identical to what it was before John had made any measurement,
Eugene must fully expect that if he then asks John about the result of his
measurement, John will say he has not yet performed any measurement. He must
further fully expect that his own examination at, and at any time after,
$t_{g}$ of \textit{D}, John's notebook, and anything else inside the
laboratory will reveal no record of any such measurement ever having been
made. Eugene's action at $t_{f}$ has, by $t_{g}$, erased all traces of John's
measurement and its result: Indeed, Eugene has succeeded in erasing all traces
of everything that happened inside the laboratory between $t_{i}$ and $t_{f}$.
Once again, he has not changed the past. But there has been a wholesale loss
of history, understood as reliable information about the past. This point will
be pursued in the concluding discussion. The present section has shown how the
pragmatist view of Healey (2012) maintains the objectivity of physical
description in the Wigner's friend scenario.

\section{Objectivity secured}

According to Healey (2012), John inside the laboratory and Eugene outside do
not assign inconsistent truth-values to claim\textit{\ C}. Moreover, each
assigns the same rich content to \textit{C} based on the decoherence of the
state of \textit{s}+\textit{D} induced by environmental interactions inside
the laboratory. This certifies the objectivity of \textit{C}'s content in the
Wigner's friend scenario. The certification depends on the fact that both John
and Eugene apply essentially the same model of decoherence to the same initial
quantum state of \textit{s}+\textit{D}.

But, according to Healey (2012), while quantum states are objective they are
also relational: the quantum state an agent should assign to a system
generally depends on the (actual or hypothetical) physical situation of that
agent. If differently situated agents assign different quantum states to a
system, this raises the possibility that their models of its decoherence will
so differ as to lead each to assign a different content to certain claims
concerning it. Relativization of quantum state ascriptions threatens
relativization of the content of NQMCs like \textit{C}. This would undermine
their objectivity in the fourth respect pointed out in section 2.

Now even though John and Eugene are differently situated in the Wigner's
friend scenario, each can choose to adopt the perspective of the other's
situation for the purpose of applying quantum theory. Eugene should assign the
same initial state as John to \textit{s} because each knows that this is the
state on which John will perform his measurement. Eugene and John are each in
the same state of ignorance as to the initial state of \textit{D} and its
(lack of) correlations with that of \textit{s}, so there is no reason for them
to assign different states either to \textit{D} alone or to \textit{s}%
+\textit{D}. (Their conclusion---that interactions within the laboratory
robustly decohere the state of \textit{s}+\textit{D} and so endow \textit{C}
with rich content---is not sensitive to fine details of the initial state of
\textit{s}+\textit{D}.)

One might object that Eugene's physical situation \textit{requires} him to
assign the state $\left\vert \psi_{E}\right\rangle $ to the contents of the
laboratory, and that since the phase of $\left\vert \psi_{E}\right\rangle $ is
not delocalized into its environment Eugene should not assign a rich content
to \textit{C. }An alternative objection would be that \textit{C} has no
unambiguous content for Eugene, since he has no principled reason to base his
assessment of that content on the degree of decoherence of the "internal"
state of \textit{s}+\textit{D}\ rather than that of $\left\vert \psi
_{E}\right\rangle $.

But in assessing the content of an NQMC such as \textit{C} about a system
(\textit{D} in this case)\textit{,}\ an agent like Eugene should base his
model of decoherence on quantum state assignments incorporating everything he
is in a position to know about the physical situation of \textit{D}, as
represented in non-quantum claims. We have assumed that Eugene is in a
position to know that John will initiate a certain interaction between
\textit{s} and \textit{D} inside the laboratory. Eugene's physical situation
does not require him to assign the state $\left\vert \psi_{E}\right\rangle $
to the contents of the laboratory when assessing the content of \textit{C}: to
do so would be to neglect information to which he has access in this situation
that is relevant to assignment of quantum states to \textit{s} and \textit{D}.

There are circumstances in which agents are so differently situated that they
correctly assign different quantum states to the same system, where each
agent's assignment is based on all information to which the agent has access,
given her physical situation.\footnote{This happens, for example, when Alice
and Bob perform spacelike separated polarization measurements on an entangled
photon pair. Knowing his outcome, Bob can assign a polarization eigenstate to
the photon entering Alice's detector. Since this information is not accessible
to Alice, she correctly assigns that photon a mixed polarization state.} If
models of decoherence based on such different assignments were to result in
these agents assigning different contents to the same NQMC, that could
threaten the objectivity of description in quantum theory. To assess the
threat we need to specify how far different agents' state assignments may
differ, and how this affects the models of decoherence they should employ in
assessing the content of relevant NQMCs.

The example in footnote 6 prompts the following restriction on different
agents' state assignments: If Alice assigns a mixed state to a system while
Bob assigns a pure state, then Bob's state vector lies in the support of
Alice's density matrix. An argument for a generalization of this restriction
has been offered (Brun \textit{et al}. 2002): Different agents' state
assignments by several density matrices are mutually compatible if and only if
the supports of all of them have a least one vector in common.

Suppose differently situated agents (such as Alice and Bob in the example)
assign quantum states to a system \textit{S} that differ in some way
consistent with this generalized restriction. Assuming this is the only
difference between the models of decoherence they use in assessing the content
of NQMCs about \textit{S}, there are no grounds for thinking they will arrive
at different assessments. By assumption, all their models share the generic
features of applying the same Hamiltonians to the same initial state of
\textit{S}'s environment with the same limitations on prior system/environment
entanglement. Since a model with these generic features will decohere the
phase of a generic pure state of \textit{S} to the same extent in the same
"basis"\footnote{The scare quotes mark the need to allow for models in which
decoherence effectively "diagonalizes" \textit{s}'s density operator in an
overcomplete basis of narrow Gaussian states.}, that is how it will model the
decoherence of every vector in the Hilbert space in which lies the vector
common to the supports of all the states assigned by differently situated
agents subject to the generalized restriction. It follows that the model of
decoherence each of these agents uses to assess the content of NQMCs about
\textit{S} will lead each agent to the same assessment of the contents of
NQMCs about \textit{S}. This means they will all agree on the content of these
NQMCs, thereby securing their objectivity.

\section{Independent Verifiability}

It is ironic that observation poses a threat to the objectivity of physical
description in quantum theory, since observation is generally used to
\textit{settle} questions about objectivity in science and daily life.
Doubting the objective presence of the dagger he saw, Macbeth tried to grasp
it: to prove that Banquo's ghost occupied his own place at table, he pleaded
his guests to see for themselves. Classical physics permits multiple,
independent observations on a system to verify a claim about its state, since
none of these need irremediably disturb that state.

But suppose an individual system is in a wholly unknown quantum state. No
single observation on it can reliably disclose its state. Repeated
observations are no better, since observing a system typically irreparably
disturbs its quantum state. So even if a system's wholly unknown quantum state
could be ascertained by a single observation, this finding could not be
checked in further observations, either by the original agent or by others. A
wholly unknown quantum state of an individual system is not as objective as a
corresponding classical state in the sense that it is not independently
verifiable. Ollivier \textit{et al}. (2004) put the point like this:

\begin{quote}
The key feature distinguishing the classical realm from the quantum substrate
is its objective existence.
\end{quote}

They propose what they call an operational definition of objectivity for a
property of a quantum system, according to which such a property is
simultaneously accessible to many observers who are able to find out what it
is without prior knowledge and who can arrive at a consensus about it without
prior agreement. Their idea is that such a property is objective to the extent
that multiple records of it exist in separate portions of the environment, so
that \textquotedblleft observers probing fractions of the environment can act
as if the system had a state of its own---an objective
state.\textquotedblright\ They say that

\begin{quote}
The existence of an objective property requires the presence of its complete
and redundant imprint in the environment as necessary and sufficient conditions.
\end{quote}

While Ollivier \textit{et al}. (2004) never say exactly what they mean by a
property, it seems clear they would count an NQMC locating a value in $\Delta$
of a magnitude $M$ on system \textit{S} as a property assignment to \textit{S}.

The state of a system in classical physics is specified by a point in phase
space: this is equivalent to an assignment of a value to each magnitude on
that system, i.e. an assignment of a property locating that value in a unit
set. The authors' stand-in for the objective state of an open quantum system
is one of the eigenvectors of the system observable corresponding to the
pointer magnitude that is selected by environmental decoherence. A NQMC
locating a value in $\Delta$ to the pointer magnitude on system \textit{S} is
taken to assign \textit{S} an \textit{objective} property solely on the
grounds that a complete and redundant imprint is present in the environment.
There are other magnitudes on \textit{S} represented by operators each of
whose eigenvectors is "close" to an eigenvector of the pointer observable. A
NQMC locating a value in $\Delta$ to such a magnitude on system \textit{S}
also counts as assigning \textit{S} a reasonably objective property because of
a complete and (slightly less) redundant imprint in the environment. Their
idea seems to be that the proliferation of imprints of properties of a quantum
system in its environment progressively objectifies its properties until these
come to mimic properties that characterize a classical state. An eigenvector
of the pointer observable stands in for an objective state by specifying what
properties of the system are objectified by the environmental interactions to
which it is subjected.

\begin{quote}
...amplification of a preferred observable happens almost as inevitably as
decoherence, and leads to objective classical reality.
\end{quote}

This is Quantum Darwinism: "the idea that the perceived \textit{classical
reality} is a consequence of the selective proliferation of information about
the system". It is not an account of classical reality, or of the actual
objective state of a quantum system. It is an account of how independent acts
of observation on a system's environment can produce consensus on properties
of a quantum system irrespective of whether or not that system has such
properties. One is reminded of Wittgenstein's remark in his
\textit{Philosophical Investigations}

\begin{quote}
As if someone were to buy several copies of the morning paper to assure
himself that what it said was true.
\end{quote}

Wittgenstein's avid reader is actually in better shape than multiple quantum
observers. The morning paper may have correctly reported what happened. But if
no NQMCs about a quantum system are true then it lacks the properties
observers attribute to it, so whatever proliferates is in fact
\textit{misinformation}. It is important not to be misled by the causal
language of imprints into thinking that objective properties of the
environment are caused by objective properties of a decohered quantum system.
As it stands,Quantum Darwinism (Ollivier \textit{et al}. 2004) fails to
provide an adequate account of objectivity in the sense of the independent
verifiability of NQMCs because it does nothing to show how those NQMCs can be
objectively true, given the quantum states of system and environment.
Groupthink does not amount to intersubjective verification. But the pragmatist
view of Healey (2012) can help the quantum Darwinist take this crucial last step.

In that view, decoherence endows certain NCMCs with a significant content that
requires even differently situated agents to seek agreement on their
truth-values. When a system is decohered by its environment, these will
include claims about the value of its pointer magnitude. They will also
include claims about magnitudes on subsystems of the environment that are
correlated with the system's pointer magnitude. Since properties not only of
the system but also of subsystems of its environment are in this sense
objective, it makes sense to ask whether objective claims about a system can
be independently verified by observing various portions of its environment.
Quantum Darwinism may now offer illuminating answers to this question by
providing quantum models of interactions between a quantum system and its
multipartite environment. This would be a way to show how differently situated
agents can come to agree on the truth-values of significant NQMCs about a
system without disturbing its state.

\section{Conclusion}

Reflection on the paradox of Wigner's friend has persuaded some people that
quantum theory cannot be understood without careful attention to the role of
conscious human experience (Wigner 1967, D'Espagnat 2005). But a pragmatist
interpretation (Healey 2012) permits a consistent and unambiguous treatment of
the paradox without reference to consciousness. Situated agents can use
quantum theory to make objective claims about the values of magnitudes in the
physical world, not just about observations of them. This has helped us to
predict and explain an enormous variety of otherwise puzzling physical
phenomena. A true claim about the value of a physical magnitude states an
objective physical fact. Ordinarily, such facts are readily independently
verifiable. But acceptance of quantum theory requires one to countenance the
possibility that in extraordinary circumstances information about the value of
a magnitude could be irretrievably erased, making this fact no longer verifiable.

A simple magnitude claim ascribes an objective property to a\ physical system.
But the \textit{content} of this claim is a function of the environmental
context. This means that the property ascribed is not intrinsic, but
relational. In the case of quantum field theory, it implies that any claim
about particles has a well-defined content only in some contexts, while other
contexts give significance to a claim about (classical) fields. In assessing
the content of a claim, an agent should consider the nature and extent of
environmental delocalization of coherence. This content does not depend on the
(actual or hypothetical) physical situation of the agent, even when agents
base their assessments on the different quantum states appropriate to their
different physical (and therefore epistemic) situations.

Science is based on observed facts, and quantum theory is no exception.
According to Healey (2012), what makes quantum theory exceptional is what it
teaches us about the nature of these facts. Bell (2004, 41) introduced the
term 'beable' because he thought

\begin{quote}
it should be possible to say of a system not that such and such may be
\textit{observed} to be so but that such and such \textit{be} so.
\end{quote}

When making a significant claim about the value of a magnitude on a quantum
system one is saying that it be so, according to Healey (2012). Perhaps this
makes this magnitude a beable in Bell's sense. But Bell (2004 52) goes on to say

\begin{quote}
the beables ... can be described "in classical terms", because they are there.
The beables must include the settings of switches and knobs on experimental
equipment, the currents in coils, and the readings of instruments.
\end{quote}

and (Bell 2004, 174)

\begin{quote}
The beables of the theory are those elements which might correspond to
elements of reality, to things which exist.
\end{quote}

These passages at least suggest that acceptance of quantum theory in no way
modifies the content of claims about values of magnitudes---a content that is
somehow established by a fixed representation relation that obtains between
language and the world (actual, or merely possible if a beable be not)!
A\ pragmatist cannot accept such a representational account of how content
accrues to a claim (Brandom 1994, 2000). According to the pragmatist
interpretation of Healey (2012), to understand quantum theory one needs
instead to adopt an alternative inferentialist account of what gives a claim
content. By modifying the inferential relations between magnitude claims,
quantum theory affects their content, rendering this contextual. By making the
content of a magnitude claim about a system a function of the environment,
acceptance of quantum theory cautions one against taking that claim to
attribute an intrinsic property to a non-contextually existing object, even
while insisting on the objectivity of the claim. This should make one think
differently even about the content of everyday claims about ordinary things
like the settings of switches and knobs on experimental equipment, the
currents in coils, and the readings of instruments.

\end{document}